\documentclass[11pt]{article}
\usepackage{moriond,epsfig}

\newcommand{\BQ}{\begin{equation}}
\newcommand{\EQ}{\end{equation}}
\newcommand{\BQA}{\begin{eqnarray}}
\newcommand{\EQA}{\end{eqnarray}}
\newcommand{\be}{\begin{eqnarray}}
\newcommand{\ee}{\end{eqnarray}}
\newcommand{\NN}{\nonumber \\}
\newcommand{\del}{\partial}

\newcommand{\x}{x_\perp}
\newcommand{\y}{y_\perp}

\newcommand{\rr}{r_\perp}

\newcommand{\beq}{\begin{eqnarray}}

\newcommand{\eeq}{\end{eqnarray}}

\newcommand{\AmS}{{\protect\the\textfont2  
  A\kern-.1667em\lower.5ex\hbox{M}\kern-.125emS}}
\def\simge{\mathrel{%
   \rlap{\raise 0.511ex \hbox{$>$}}{\lower 0.511ex \hbox{$\sim$}}}}
\def\simle{\mathrel{
   \rlap{\raise 0.511ex \hbox{$<$}}{\lower 0.511ex \hbox{$\sim$}}}}

\begin{document}
\vspace*{4cm}
\title{UNDERSTANDING GEOMETRIC SCALING AT SMALL X 
}

\author{E.~IANCU$^a$, \underline{K.~ITAKURA}$^b$, and L.~McLERRAN$^c$}
\address{$^a$ Service de Physique Theorique, CE Saclay, F-91191 
        Gif-sur-Yvette, France\\
$^b$ RIKEN BNL Research Center, BNL, Upton, NY 11973, USA\\
$^c$ Nuclear Theory Group, Brookhaven National Laboratory,
        Upton, NY 11973, USA}

\maketitle\abstracts{
Geometric scaling is a novel scaling phenomenon observed in
deep inelastic scattering at small~$x$: the total $\gamma^* p$
cross section depends upon the two kinematical variables $Q^2$ 
and $x$  only via their combination $\xi\equiv Q^2 R_0^2(x)$, 
with $R_0^2(x) \propto x^{\lambda}$. At sufficiently low $Q^2$, 
below the saturation scale $Q_s^2(x)$ ($\sim$  a few GeV$^2$),
this phenomenon finds a natural explanation as a property 
of the Color Glass Condensate, the high-density matter made
of saturated gluons.
To explain the
experimental observation of geometric scaling up to much higher
values of $Q^2$, of the order of 100 GeV$^2$, we study the solution
to the BFKL equation subjected to a saturation boundary condition
at $Q^2\sim Q_s^2(x)$. We find that the scaling extends indeed
above the saturation scale, within a window  
$1 \simle \ln(Q^2/Q_s^2) \ll \ln(Q_s^2/\Lambda^2_{\rm QCD})$,
which is consistent with phenomenology.}

\section{Geometric Scaling at Small $x$}
\vspace{-2mm}
Recently, Sta\'sto, Golec-Biernat and
Kwieci\'nski have shown~\cite{geometric} that the HERA data on deep 
inelastic scattering 
at low $x$, which are a priori functions of two independent variables
--- the photon virtuality $Q^2$ and the Bjorken variable $x$ ---, 
are consistent with scaling in terms of the variable
\vspace{-2mm}
\begin{equation}\label{scaling_variable}
	\xi = Q^2 R^2_0(x)
\end{equation}
where $R^2_0(x) = (x/x_0)^{\lambda}/Q_0^2$ with the parameters 
$\lambda = 0.3 \div 0.4$,  $Q_0 = 1 ~{\rm GeV}$, and
$x_0 \sim 3 \times 10^{-4}$ determined to fit the data. 
In particular, the data for the virtual 
photon total cross section at $x < 0.01$ and $0.045 < Q^2 < 450~{\rm GeV}^2$
are consistent with being only a function of $\xi$. 
This is the phenomenon called the ``geometric scaling''.
In this talk, we argue that this is a manifestation of gluon saturation 
and explain why it holds even outside the saturation regime. 
For more details, see Ref.~2.

\section{Geometric Scaling from the Color Glass Condensate}
\vspace{-2mm}

At very high energies, the cross-sections for hadronic
processes
are dominated by the small-$x$ gluons in the hadron
wavefunction. 
These gluons make a high density matter which is believed to
reach a 
saturation regime, and become a Color Glass Condensate 
(See Ref.~3 for a recent review and more references.)
This picture provides us with a natural interpretation 
of the geometric scaling. Indeed, in the saturation regime, 
there is only one intrinsic scale 
in the problem, the saturation momentum $Q_s(x)$ itself.
($1/Q_s(x)$ is the typical transverse size of the saturated
gluons.)
Thus, all physical quantities should be expressed as a
dimensionless 
function of $Q^2/Q_s^2(x)$ times some power of $Q_s^2(x)$
giving the right 
dimension. This, together with the fact that $Q_s^2(x)$
increases
as a power of the energy: $Q_s^2(x) \sim Q_0^2\,
x^{-\lambda}$
(this follows from the general equations~\cite{Cargese}
 describing the  evolution
 of the Color Glass Condensate with $x$; see also
below), suggests
the identification between $Q_s^2(x)$ and the function
$1/R_0^2(x)$ in 
the scaling variable (\ref{scaling_variable}).

We shall briefly show how to determine the energy 
dependence of $Q_s$ by using non-linear evolution equation. 
At small $x$, the total $\gamma^* p$ cross section 
$\sigma_{\rm total}^{\gamma^*p}$
can be calculated as follows ($\sigma_{\rm total}^{\gamma^*p}=\sigma_{\rm T}(\tau, Q^2)+\sigma_{\rm L}(\tau, Q^2), \ 
\tau\equiv \ln(1/x)$) 
: 
\be\label{sigmaDIS}
\sigma_{\rm T,L}(\tau,Q^2)\,=\,
\int_0^1 dz \int d^2r_\perp\,|\Psi_{\rm T,L}(z,r_\perp;Q^2)|^2\,
\sigma_{\rm dipole}(\tau,r_\perp),
\ee
where 
$\Psi_{\rm T,L}(z,r_\perp;Q^2)$ is the light-cone wavefunction of the virtual 
photon splitting into a $q\bar q$ pair (the ``dipole'') with 
transverse size $r_\perp$ and a fraction $z$ of the photon's longitudinal 
momentum carried by the quark, and $\sigma_{\rm dipole}(\tau,r_\perp)$ 
is the dipole-proton cross-section given by 
\be\label{sigmadipole}
\sigma_{\rm dipole}(\tau,r_\perp)\,=\,2\int d^2b_\perp\,
(1-S_\tau(x_{\perp},y_{\perp})),\quad\  
S_\tau(x_{\perp},y_{\perp})\equiv \frac{1}{N_c}\,
\left\langle {\rm tr}\big(V^\dagger(x_{\perp}) V(y_{\perp})\big)
\right\rangle_{\tau},\ee
with $r_\perp=x_{\perp}-y_{\perp}$ and
the impact parameter $b_\perp=(x_{\perp}+y_{\perp})/2$ (the
quark is at $x_{\perp}$, and the antiquark at $y_{\perp}$).
In eq.~(\ref{sigmadipole}), $V^\dag$ (or $V$) is 
a path ordered exponential of the color field created 
in the proton by color source at rapidities $\tau'<\tau$.
The brackets in the definition of the 
$S$-matrix element $S_\tau(x_{\perp},y_{\perp})$ 
refer to the average over all configurations of these color sources.
One can check that if $\sigma_{\rm dipole}$ shows the scaling property
$\sigma_{\rm dipole}(\tau,r_\perp)=\sigma_{\rm dipole}(r_\perp^2Q_s^2(\tau))$,
it is transmitted to the total cross-section.
Furthermore, since we assume transverse homogeneity of the proton, 
the integral over the impact parameter gives a trivial contribution, 
which allows us to relate the scaling property of the $S$-matrix elements 
$S_\tau(\x,\y)=S_\tau(r_\perp)$ to that of $\sigma_{\rm dipole}$.  

The scattering matrix $S_\tau(r_\perp)$ is determined by solving the 
non-linear evolution equation called the Balitsky-Kovchegov (BK) equation
(below,  $\bar \alpha_s= N_c\alpha_s/\pi$) :
\beq
{\partial \over {\partial \tau}} S_\tau( \x\!\! -\y)\! =\! 
-\bar \alpha_s\!\! \int\! {{d^2z_\perp} \over {2\pi}} 
{{(x_\perp\!\!-\y)^2} \over {(x_\perp\!\!-z_\perp)^2 (\y\!\!-z_\perp)^2}} 
 \Big(S_\tau(x_\perp\!\!-\y)- S_\tau(x_\perp\!\!-z_\perp)
S_\tau(z_\perp-\y)   \Big).\label{BK}
\eeq
We can approximately solve \cite{IIM} 
this equation at large transverse distance
(corresponding to momenta $Q^2\ll Q^2_s(x)$), and the solution indeed 
shows the scaling property $S_\tau(\rr) =f(\rr^2 Q_s^2(\tau))$.
Furthermore, if one assumes the scaling in the BK equation, it reduces to an 
equation for the saturation scale: 
$
\frac{\del}{\del \tau}\ln Q_s^2(\tau) = c\, \bar\alpha_s, 
$
where $c$ is a constant to be determined later.   
Therefore, the $\tau$ dependence of $Q_s$ has been 
determined as
\BQ
Q_s^2(\tau)= \Lambda^2 \,{\rm e}^{c\bar\alpha_s \tau},
\label{saturation_scale_fixed}
\EQ
with $\Lambda$ fixed by the initial condition (typically, 
$\Lambda\sim \Lambda_{\rm QCD}$).

Hence, we have seen that the Color Glass Condensate naturally leads to
the geometric scaling. 
However, the fundamental problem with this argument is that it is valid 
only for $Q^2$ less than or of the order of the saturation momentum, 
which is at most several ${\rm GeV}^2$, while the fit of 
Ref.~1 extends up to $Q^2$ of the order of several hundred GeV$^2$.

\section{Geometric Scaling above the Saturation Scale}

\vspace{-2mm}
To understand the reason why the actual scaling region is much larger 
than the saturation region, we shall investigate the BFKL equation, 
which is the appropriate evolution equation above the saturation scale:
The dipole-hadron scattering amplitude 
${\cal N}_\tau(\rr)\equiv 1 - S_\tau( \rr)$ is small ${\cal N}_\tau(\rr)\ll 1$
for a small dipole, $\rr\ll 1/Q_s(\tau)$,
and one can linearize eq.~(\ref{BK}) with respect to 
${\cal N}_\tau(\rr)$, which yields the BFKL equation.
The solution to the BFKL equation  is expressed as 
the Mellin integral with respect to the transverse coordinates: 
\vspace{-2mm}
\BQ\label{Mellin}
{\cal N}_\tau(\rr)=\int_{C} \frac{d\lambda}{2\pi i} 
\left(\frac{\rr^2}{\ell^2}\right)^\lambda  
{\rm e}^{\bar\alpha_s \tau
   \left\{2\psi(1)-\psi(\lambda)-\psi(1-\lambda)\right\}}
\EQ
where $\ell^2=1/\Lambda^2$ and 
$\psi(\lambda)=\Gamma'(\lambda)/\Gamma(\lambda)$.
We can perform the saddle point approximation 
in the kinematical region of interest: $\bar\alpha_s \tau \gg 1$ 
and $r=\ln Q^2 /\Lambda^2 =-\ln \rr^2/\ell^2 \gg 1.$
The position of the saddle point depends on the ratio $r/\bar\alpha_s\tau$, 
and we find two limiting cases. 
(i) When $r/\bar\alpha_s\tau$ is large: 
the saddle point is close to $\lambda=1$.
(ii) When $r/\bar\alpha_s\tau$ is small: 
the saddle point is close to $\lambda=1/2$.
The solution (\ref{Mellin}) in case (i) corresponds 
to the double log approximation (DLA) of the DGLAP solution. 
Since the saddle point is given by 
$\lambda\simeq 1- \sqrt{\bar\alpha_s\tau/r}$, this case is realized when 
$\sqrt{\bar\alpha_s\tau/r}\ll 1/4,$ namely, 
$16\bar\alpha_s\tau \ll \ln Q^2/\Lambda^2$. 
On the other hand, case (ii) yields the standard BFKL solution 
(we ignored fluctuations around the saddle point):
\vspace{-2mm}
\BQ\label{solution_cBFKL}
{\cal N}_\tau(\rr)\simeq \sqrt{\frac{\rr^2}{\ell^2}}\, 
{\rm e}^{\omega\bar\alpha_s\tau}
\exp \left\{ \frac{- \ln^2 \left({\rr^2/ \ell^2}\right) }
                  {2\beta \bar\alpha_s\tau} 
\right\},
\EQ
where $ \beta 
=28\zeta(3)=33.67$ and
$\omega 
=4\ln 2=2.77$.
Since the saddle point is estimated as $\lambda\simeq 
1/2 + r/\beta\bar\alpha_s\tau$,
this case is realized when  $r/\beta\bar\alpha_s\tau\ll 1/4$, namely, 
$\ln Q^2/\Lambda^2 \ll 8\bar\alpha_s\tau$.
Therefore, we can roughly divide the whole kinematical region into 
three parts (see Fig.~1): 
(i) the DLA regime, $16\bar\alpha_s\tau \ll \ln Q^2/\Lambda^2$, 
(ii) the BFKL regime, $c\bar\alpha_s\tau \ll \ln Q^2/\Lambda^2 \ll 
8\bar\alpha_s\tau$, where the lower bound comes from the fact that 
the linear BFKL equation is just an approximation of the non-linear 
BK equation and thus valid only above the saturation scale 
$Q_s^2(\tau)\propto{\rm e}^{c\bar\alpha_s\tau}$. 
So, the last regime is (iii) Color Glass Condensate, 
$\ln Q^2/\Lambda^2 \ll c\bar\alpha_s\tau$.
Here the number $c=4\div 5$ is determined below. 
In the following, we discuss only the BFKL solution (\ref{solution_cBFKL}) 
because we are interested in the region just above the saturation scale.

As already mentioned, the BFKL solution (\ref{solution_cBFKL})
cannot be used below the saturation scale where ${\cal N}_\tau(\rr)\simeq 1$.
But we can still use this equation to determine the
saturation momentum from the condition (the saturation
criterion):
\vspace{-2mm}
\BQ\label{criterion}
{\cal N}_\tau(\rr=1/Q_s(\tau))\,=\,1.
\EQ
Physically, this is the matching condition of the BFKL solution at 
the saturation scale. By solving eq.~(\ref{criterion}), one obtains 
a result consistent with eq.~(\ref{saturation_scale_fixed}) 
with the coefficient $c$  determined to be $c= 4\div 5$. 
To see the behavior of the BFKL solution near the saturation momentum, 
we expand the exponent of the solution ($\ln {\cal N}_\tau(\rr)$) 
around the saturation momentum with respect to $\ln Q^2/\Lambda^2 \ (=
-\ln \rr^2/\ell^2)$:
\vspace{-2mm}
\BQA
{\cal N}_\tau(\rr=1/Q)
&=& \exp\left\{\ln {\cal N}_\tau(Q_s) + 
\left.\frac{\del\ln {\cal N}_\tau(Q)}{\del \ln Q^2}\right\vert_{Q_s^2}
\left(\ln {Q^2\over \Lambda^2} - \ln {Q_s^2\over \Lambda^2}\right)+\cdots 
\right\}\NN
&\simeq& \left(\frac{Q^2}{Q_s^2(\tau)}\right)^{-\lambda_s},\quad 
-\lambda_s\equiv \left.\frac{\del\ln {\cal N}_\tau(Q)}{\del \ln Q^2}
\right\vert_{Q_s^2(\tau)},
\EQA
where we have taken up to the first order of the expansion. 
The zeroth order vanishes due to the saturation criterion (\ref{criterion}).
What is remarkable in the last result is that it shows the geometric scaling 
because the power $\lambda_s$ is just a number and independent of $\tau$. 
This is a consequence of the particular property of the evolution equation,
the scale-invariance of the BFKL kernel in transverse space.  
Keeping just the first nonzero term in this expansion is a good
approximation as long as
\BQ
\left(\ln\frac{Q^2}{\Lambda^2}-\ln \frac{Q_s^2}{\Lambda^2}\right) 
= \ln\frac{Q^2}{Q_s^2}\, \ll\,  \ln\frac{Q_s^2}{\Lambda^2}=c\bar\alpha_s\tau.
\EQ
Therefore, the geometric scaling approximately holds even above the 
saturation scale if the scale satisfies the above inequality. 
For $Q_s \sim 1\div 2~{\rm GeV}$ and $\Lambda\sim
200~$MeV, the upper scale on $Q^2$ is $Q_s^4/\Lambda^2
\sim 25 \div 400~ {\rm GeV}^2$. 
Furthermore, it should be noticed that the extended scaling region 
$\ln Q^2/\Lambda^2 - c\bar\alpha_s\tau \ll c\bar\alpha_s\tau$ 
almost coincides with the domain of validity for the BFKL 
saddle-point ($\ln Q^2/\Lambda^2 \ll 8\bar\alpha_s\tau $).  
Indeed, if one recast the BFKL solution (\ref{solution_cBFKL}) 
in a form which makes its scaling properties obvious 
\vspace{-2mm}
\be\label{BFKLscaling}
{\cal N}_\tau(\rr=1/Q)
\simeq\, \left(Q^2/Q_s^2(\tau)\right)^{-\lambda_s}
\exp \left\{ -\frac{1}{2 \beta \bar\alpha_s\tau} \left(
\ln\frac{Q^2}{ Q_s^2(\tau)}\right)^2\right\} ,
\ee
one finds 
this has the structure of the ``second-order expansion'' 
with a particularly small coefficient.  
Therefore,  we conclude that, in the whole kinematical range where 
it applies, the BFKL solution (\ref{solution_cBFKL}) is almost
a scaling solution (see Fig.~1).

\vspace{2mm}

In summary, we have shown that the geometric scaling predicted at 
low momenta $Q^2 \simle Q_s^2$ by the Color Glass Condensate
is preserved by the BFKL equation up to relatively large $Q^2$ 
momenta, within the range $1 \simle \ln(Q^2/Q_s^2) \ll 
\ln(Q_s^2/\Lambda^2_{\rm QCD})$. The matching procedure of the 
linear BFKL solution to the saturation regime at $Q^2 \simeq Q_s^2$
was essential to carry the information about saturation to the 
regime above the saturation scale.

\begin{figure}
\begin{center}
\psfig{figure=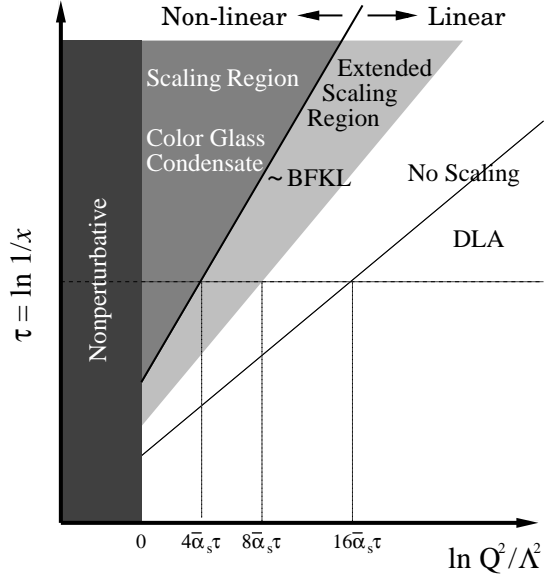,height=3in}
\caption{Domains for quantum evolution in DIS at small $x$}
\end{center}
\end{figure}

\end{document}